# Implementation of Linear Parameter Varying System to Investigate the Impact of Varying Flow Rate on the Lithium-ion Batteries Thermal Management System Performance

Pedram Rabiee[1], Mohammad Hassan Saidi[2]

[1]Sharif University of Technology; rabiee_pedram@mech.sharif.edu
[2]Sharif University of Technology; saman@sharif.edu

**Abstract**
Battery thermal management system is the indispensable part of the electric vehicles working with Lithium-ion batteries. Accordingly, lithium-ion batteries modelling, battery heat generation, and thermal management are the main focus of researchers and car manufacturers. To fulfil the need of manufacturers in the design process, a faster model than time-consuming Computational Fluid Dynamics models (CFD) is required. Reduced Order Models (ROM) address this requirement to maintain the accuracy of CFD models while could be compiled faster. Linear Time Invariant (LTI) reduced order model has been used in the literature; however, due to the limitation of LTI system, considering constant flow rate for the cooling fluid, a Linear Parameter Varying system with three scheduling parameters was developed in this study. It is shown that LPV system result could fit accurately to CFD result in conditions that LTI system cannot maintain accuracy. Moreover, it is shown that applying varying water flowrate could result in smoother temperature profile.

**Introduction**
The attention to thermal management of lithium-ion batteries has been increased in recent years. The objective of battery thermal management system is to lower maximum temperature and maximum local temperature difference in cell to decrease cell degradation. Enough attention has been paid to CFD modeling of different thermal management system designs: air cooling systems [1]–[3], liquid cooling systems [4]–[7] and phase change material (PCM) based cooling systems [8]–[10]. However, CFD models are too large and slow to be implemented in the design process. In the repetitive transient analysis of the cooling system performance under different driving cycles or charge and discharge profiles, the need for smaller models that maintain the accuracy of CFD models arises. Reduced Order Models (ROM) were used in the literature in different levels of battery modeling, from ROMs developed for the physics-based electrochemical models of the battery [11] to ROMs that predict the thermal behavior of the battery pack [12]–[14]. The implementation of ROMs for the thermal management system prepares the model for the integration with the vehicle model. This integration is essential for studying the battery thermal management performance under real driving conditions, and battery thermal condition impact on the vehicle performance. Hu et al. [13] introduced the implementation of Linear Time Invariant (LTI) ROM for battery thermal management problem. It was shown in this study that LTI systems can calculate the average temperature of the battery cell under arbitrary heat generation based on the step-responses extracted from the CFD model, and the result was compared with the CFD result. However, as it is to be shown in this study, LTI system can predict inaccurate results under heat generations that differ highly from the heat generation that was used in the step-response extraction process. Thus, a Linear Parameter Varying (LPV) system was developed with heat generation as one of its scheduling parameter to obtain more accurate results than LTI's result. Furthermore, another limitation of LTI systems is that as the step-responses required to be extracted under constant flow rate, a single LTI system cannot predict the system behavior under varying flow rate. Hence, an LPV system with three scheduling parameters –inlet water flow rate, inlet water temperature, and heat generation– was developed to investigate the impact of varying flow rate on thermal management system performance.

**Model and Methodology**
The electrical behavior of the battery cell is modeled with Equivalent Circuit Model (ECM), based on the experimental data for the LG P1.4 (14.8) battery cell released in [15]. The proposed model shown in "Figure 1" mixes the merits of both Chen's –having state-of-charge (SOC) estimation circuit– [16] and Plett's –temperature dependent model parameters– [15] ECM.

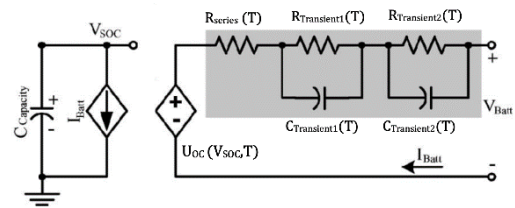

Figure 1. Proposed Equivalent Circuit Model

Heat generation in cell is calculated using Bernardi [17] equation:

$$\dot{q} = I(U_{OC}-V) - I\left(T\frac{\partial U_{OC}}{\partial T}\right) \tag{1}$$

where, $U_{OC}$ is the open-circuit voltage, $V$ is the cell voltage, and $I$ is the current density. The term $\frac{\partial U_{OC}}{\partial T}$ can be obtained from experimental data of the battery cell. Thomas et al. [18] result as shown in "Figure 2" was used to obtain this term. The ECM model developed in ANSYS Simplorer as shown in "Figure 3".

Step-responses for the extraction of LTI systems extracted from CFD model in ANSYS Fluent. "Figure 4" shows the battery cell and cooling plates design.

The model was meshed in ANSYS Meshing and the numerical calculation was conducted in ANSYS Fluent 18.0. Laminar model was employed as the maximum Reynolds number was less than 200 in all cases. Insulation boundary condition was taken for side surfaces. Independent test of grid number was carried out to ensure accuracy *of* the calculation. The result is shown in "Figure 5".

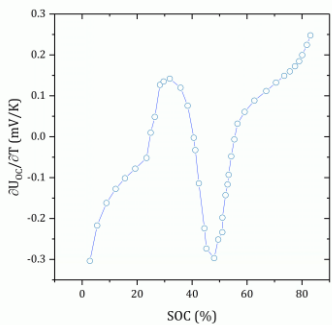

Figure 2. $\frac{\partial U_{OC}}{\partial T}$ as a function of SOC [18].

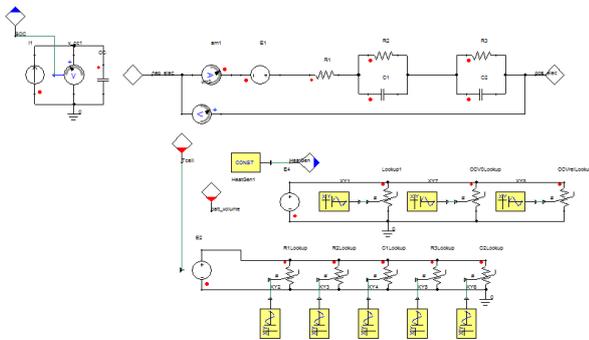

Figure 3. Equivalent circuit model development in ANSYS Simplorer.

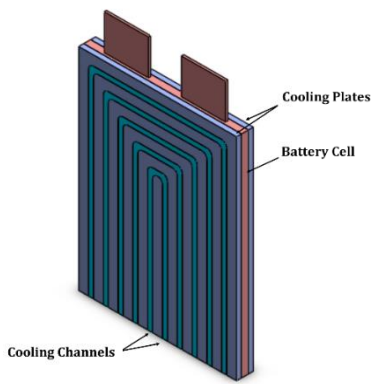

Figure 4. Battery cell surrounded by two cooling plates.

The thermodynamic properties of battery, water, and cooling plate is listed in "Table 1." Cooling plate is made of aluminum.

Table 1. Thermodynamic properties of battery, water, and cooling plates [5].

| Material | Aluminum | Water | Battery |
|---|---|---|---|
| $\rho$ (kg m$^{-3}$) | 2719 | 998.2 | 2500 |
| $C_p$ (J kg$^{-1}$ K$^{-1}$) | 871 | 4128 | 100 |
| $\lambda$ (W m$^{-1}$ K$^{-1}$) | 202.4 | 0.6 | 3 |
| $\mu$ (Pa.s) | - | 1.003×10$^{-3}$ | - |

ISME2018, 24-26 April, 2018

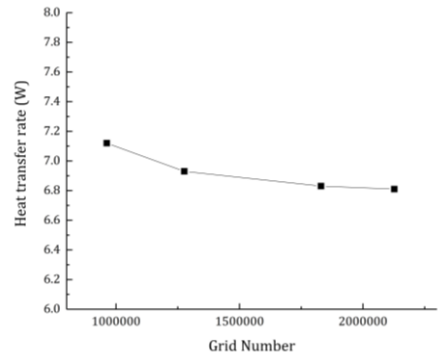

Figure 5. Independent test of grid number.

"Figure 8" shows the LPV grid with three scheduling parameters. The LPV grid values for the scheduling parameters is listed in "Table 2."

Table 2. Grid values for the scheduling parameters.

| $\dot{q}$ (W/m$^3$) | $\dot{m}$ (kg/s) | $T_i$ (°C) |
|---|---|---|
| 1×10$^5$ | 2×10$^{-4}$ | 5 |
| 5×10$^5$ | 5×10$^{-4}$ | 10 |
| 5×10$^6$ | 1×10$^{-3}$ | 15 |
| | 2×10$^{-3}$ | 20 |
| | 3×10$^{-3}$ | |

The LPV system was developed by co-simulation between ANSYS Simplorer and Simulink. The ECM model in Simplorer was coupled with LPV model in Simulink as shown in "Figure 6" and "Figure 7." The state-space models created from step-responses for different grid values were imported in MATLAB workspace to be used by LPV system block.

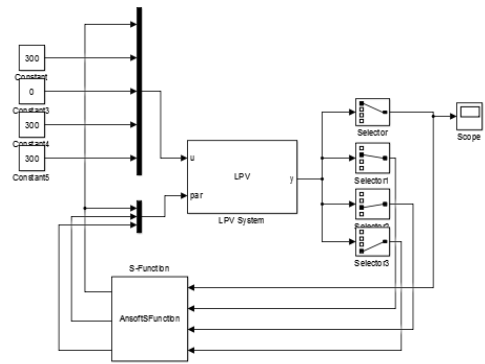

Figure 6. LPV system with 3 scheduling parameters co-simulation with Simplorer battery model.

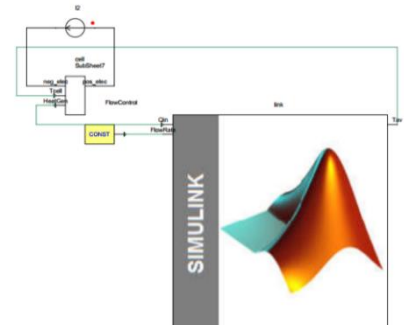

Figure 7. Simplorer battery model in co-simulation with Simulink LPV model.

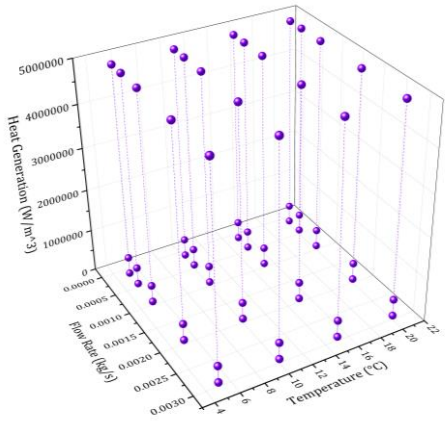

Figure 8. LPV grid.

**Results and discussions**

A heat generation as shown in "Figure 9" was considered to compare LTI and CFD results. The inlet flow rate and inlet water temperature for this comparison was taken to be 0.002 kg/s and 5°C, respectively.

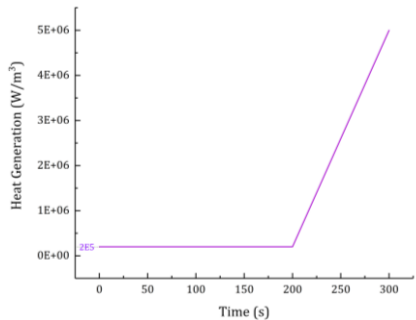

Figure 9. Heat generation used for LTI validation.

The result is shown in "Figure 10" for different LTI models created based on step-responses extracted under different heat generations. As it can be seen, LTI model results do not match perfectly to CFD result, and the results are different from what Hu et al. [13] predicted about LTI systems. This is due to the fact that in this simulation the chosen flow rate is high (0.002 kg/s) and the inlet water temperature is low (5°C); thus, for t<200s, as the heat generation is respectively low, the temperature of the cell will be decreased, which can be seen in the CFD result. However, for LTI systems created with high heat generation in the step-response extraction process (i.e. $\dot{q}=5\times10^6$ W/m$^3$), this decrease was not monitored, as temperature merely increased with high level of heat generation in the step-response. Accordingly, when low heat generation (i.e. $\dot{q}=2\times10^5$ W/m$^3$) was imported as input to these LTI systems, these systems predicted a subtle amount of temperature increase in spite of temperature decrease. If the system had only experienced the temperature increase from its initial condition of T=300 K, the results would have changed extensively. This temperature increase from initial condition is the case presented in [13]. Actually, a heat generation such as what is presented in "Figure 9" is not experienced by the cell in actual working condition, and if so, the results would match the CFD results as [13] presented. However, for obtaining perfect match between LTI and CFD results for this case, an LPV system was developed using co-simulation between Simplorer and Simulink. State-space extracted models were introduced in MATLAB workspace to be used in LPV system block in the Simulink environment. The scheduling parameter for this LPV system is heat generation, which is also the input of the system. LPV grid was formed based on 7 heat generation step-responses. In "Figure 3" the heat generation used for step-response extraction is listed.

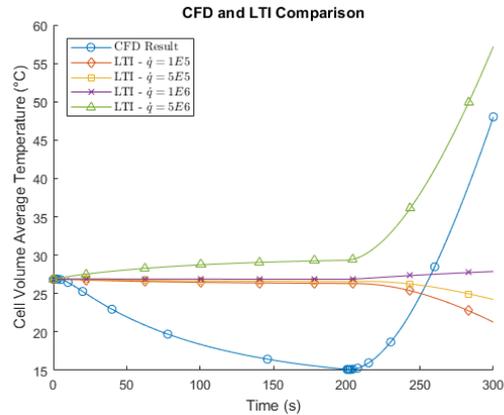

Figure 10. CFD and LTI result comparison.

Table 3. Heat generation used for step-response extraction.

| $\dot{q}\left(\frac{W}{m^3}\right)$ | $8\times10^4$ | $1\times10^5$ | $5\times10^5$ | $1\times10^6$ | $5\times10^6$ | $1\times10^7$ | $5\times10^7$ |
|---|---|---|---|---|---|---|---|

The LPV result is compared with CFD result as shown in "Figure 11." It can be seen that the error is under 4% for the entire time period.

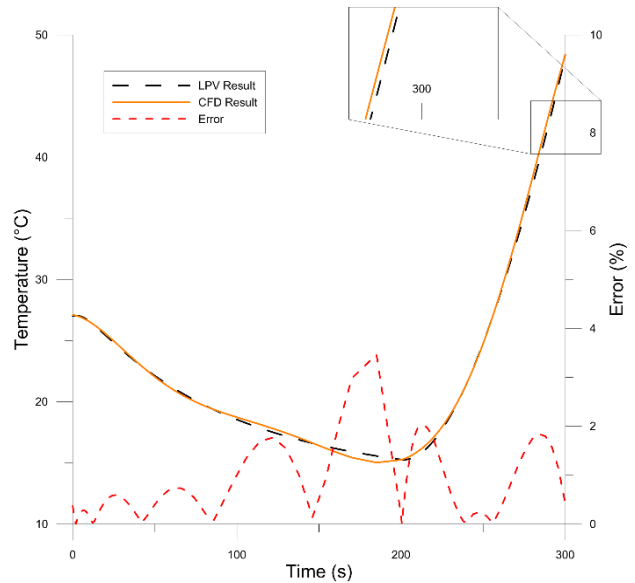

Figure 11. LPV and CFD result comparison.

As it is stated earlier, for the analysis of a system under varying flow rate, LPV system is required to be implemented; hence, with the aid of LPV system with flow rate as one of its scheduling parameters, the impact of varying flow rate on cell average temperature was studied. A varying non-linear heat generation was used for this work as shown in "Figure 12." Four flow rate profile were considered as shown in "Figure 13." The flow rate for cases 2-4 were chosen to be proportional to heat generation profile. The average flow rates were equal in all these cases and it is 0.0008 kg/s but the standard deviation of them are different. "Table 4" lists the coefficient of variation for these cases. "Case 4" has the most erratic profile. The results are shown in "Figure 14." It can be seen that proportional flow rate to heat generation can decrease the standard deviation in the average temperature, and it smooths the temperature profile.

Table 4. Coefficient of variation of flow rate in the proposed cases

|  | Case 1 | Case 2 | Case 3 | Case 4 |
|---|---|---|---|---|
| $\dfrac{\text{STD}}{\text{Mean}}$ (%) | 0 | 2.8 | 8.4 | 14 |

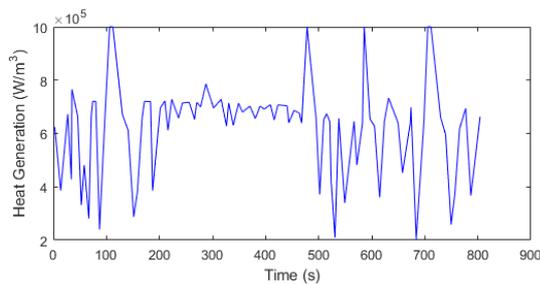

Figure 12. Heat generation used for varying flow rate effect investigation.

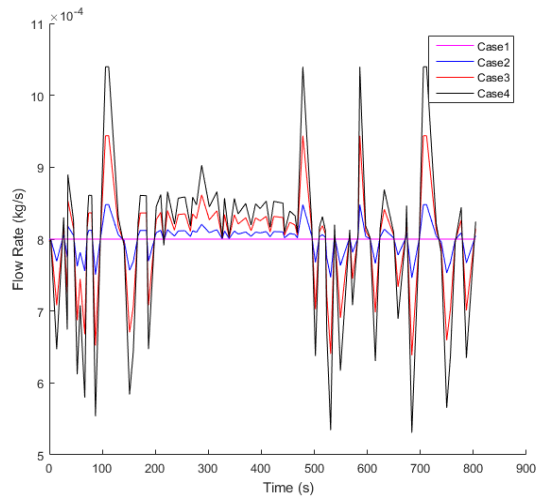

Figure 13. Flow rate profiles used for the study of varying flow rate effect investigation.

## Conclusions

The effect of implementation of Linear Parameter Varying system was investigated in this study. It was shown that Linear Time Invariant systems could not maintain accuracy under circumstances in which the heat generation in the cell is highly different from the heat generation used in the step-response extraction process. Linear Parameter Varying system with heat generation as one of its scheduling parameters, however, could retain accuracy. Furthermore, Linear Time Invariant systems cannot predict the system behavior under varying flow rate or inlet water temperature. Hence, a Linear Parameter Varying system with three scheduling parameters was developed ant it was shown that applying varying flow rate that is proportional to the heat generation can result in smoother average temperature, and decrease the temperature variations under erratic heat profile.

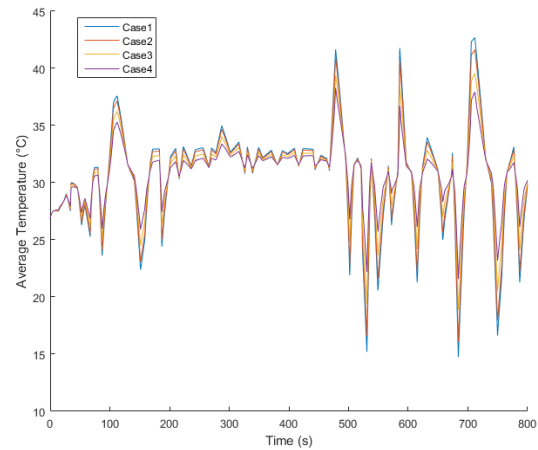

Figure 14. Average temperature of cell under 4 cases in which a flow rate proportional to heat generation was considered.

**Nomenclatures**
$C_p$    Specific heat at constant pressure
$I$      Current density
$\dot{m}$    Flow rate
$\dot{q}$    Heat generation
$T$      Temperature
$U_{OC}$  Open-circuit voltage
$V$      Cell voltage

**Greek symbols**
$\lambda$    Thermal conductivity
$\mu$    Dynamic viscosity
$\rho$   Density